# Astrobiology



**Spectral signatures of photosynthesis II: coevolution with other stars and the atmosphere on extrasolar worlds**

| | |
|---|---|
| Journal: | *Astrobiology* |
| Manuscript ID: | AST-2006-0108 |
| Manuscript Type: | Research Articles (Papers) |
| Date Submitted by the Author: | 04-Dec-2006 |
| Complete List of Authors: | Kiang, Nancy; NASA Goddard Insitute for Space Studies; CalTech, Infrared Processing and Analysis Center (IPAC)<br>Segura, Antigona; Universidad Nacional Autónoma de México, Instituto de Ciencias Nucleares<br>Tinetti, Giovanna; European Space Agency, Institut d'Astrophysique de Paris<br>Govindjee, Govindjee; University of Illinois at Urbana-Champaign<br>Blankenship, Robert; Washington University, Department of Biology and Chemistry<br>Cohen, Martin; University of California, Berkeley, Radio Astronomy Laboratory<br>Siefert, Janet; Rice University, Department of Statistics<br>Crisp, David; California Institute of Technology, NASA Jet Propulsion Laboratory<br>Meadows, Victoria; California Institute of Technology, Spitzer Science Center |
| Keyword: | Photosynthesis, Biosignatures, Red Edge, Extrasolar Terrestrial Planets, Atmospheric Compositions |
| | |

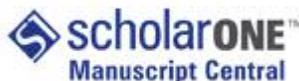





**Short title:**
**Photosynthetic spectra on extrasolar planets**

**Full title:**
**Spectral signatures of photosynthesis II:  coevolution with other stars and the atmosphere on extrasolar worlds**

*Nancy Y. Kiang[1,10], Antígona Segura[2,10], Giovanna Tinetti[3,10], Govindjee[4], Robert E. Blankenship[5], Martin Cohen[6,10], Janet Siefert[7,10], David Crisp[8,10], Victoria S. Meadows[9,10]

[1]NASA Goddard Institute for Space Studies, U.S.A.
[2]Instituto de Ciencias Nucleares, Universidad Nacional Autónoma de México
[3]Institut d'Astrophysique de Paris, European Space Agency
[4]Departments of Plant Biology and Biochemistry, University of Illinois at Urbana-Champaign, U.S.A.
[5]Department of Biology and Chemistry, Washington University, U.S.A.
[6]Radio Astronomy Laboratory, University of California, Berkeley, U.S.A.
[7]Department of Statistics, Rice University, U.S.A.
[8] NASA Jet Propulsion Laboratory, California Institute of Technology, U.S.A.
[9]Spitzer Science Center, California Institute of Technology, USA
[10]NASA Astrobiology Institute – Virtual Planetary Laboratory

*To whom correspondence should be addressed:  Nancy Y. Kiang, NASA Goddard Institute for Space Studies, New York, NY  10025, fax1: (626) 568-0673, fax2: (212) 678-5552, tel1: (626) 395-1815, tel2: (212) 678-5587, cell: (949) 439-3416, email: nkiang@giss.nasa.gov

*"Apparently the vegetable kingdom in Mars, instead of having green for a dominant colour, is of a vivid blood-red tint."  H.G. Wells, The War of the Worlds, 1898*

*Abstract.*  As photosynthesis on Earth produces the primary signatures of life that can be detected astronomically at the global scale, a strong focus of the search for extrasolar life will be photosynthesis, particularly photosynthesis that has evolved with a different parent star.  We take planetary atmospheric compositions simulated by Segura, et al. (2003, 2005) for Earth-like planets around modeled F2V, K2V, M1V, and M5V stars and around the active M4.5V star AD Leo;  our scenarios use Earth's atmospheric composition as well as very low $O_2$ content in case anoxygenic photosynthesis





DRAFT, Kiang, Astrobiology

dominates.  With a line-by-line radiative transfer model, we calculate the incident

spectral photon flux densities at the surface of the planet and under water.  We identify

bands of available photosynthetically relevant radiation and find that photosynthetic

pigments on planets around F2V stars may peak in absorbance in the blue, K2V in the

red-orange, and M stars in the NIR, in bands at 0.93-1.1 µm, 1.1-1.4 µm, 1.5-1.8 µm, and

1.8-2.5 µm.  However, underwater organisms will be restricted to wavelengths shorter

than 1.4 µm and more likely below 1.1 µm.  M star planets without oxygenic

photosynthesis will have photon fluxes above 1.6 µm curtailed by methane.  Longer-

wavelength, multi-photosystem series would reduce the quantum yield but could allow

for oxygenic photosystems at longer wavelengths. A wavelength of 1.1 µm is a possible

upper cut-off for electronic transitions vs. only vibrational energy;  however, this cut-off

is not strict, since such energetics depend on molecular configuration.  M star planets

could be a half to a tenth as productive as Earth in the visible, but exceed Earth if useful

photons extend to 1.1 µm for anoxygenic photosynthesis.  Under water, organisms would

still be able to survive UV flares from young M stars and acquire adequate light for

growth.

## 1.1.    Introduction: Planet finding missions, biosignatures of

## photosynthesis on Earth, literature review on extrasolar photosynthesis

Although we now know that Mars has no surface vegetation (we have not yet

ruled out other life), H. G. Wells was nonetheless prescient in speculating that

photosynthetic organisms on another planet might evolve to have a different dominant







# DRAFT, Kiang, Astrobiology

color than the green that is prevalent on Earth. Over 200 giant planets in other solar

systems have been discovered since 1993. In August of 2004, the discovery of three

Neptune-mass planets via the radial velocity technique (Butler, et al, 2004; McArthur, et

al., 2004) pushed the lower mass limit of discovered planets down from gas giants like

Jupiter to close to those of the smaller ice giant planets in our Solar System. The

microlensing technique has also led to the recent discovery of a 13 Earth mass planet

(Gould et al., 2006), implying that they are common. The recent discoveries of a 7.5

Earth mass planet (Rivera *et al.*, 2005) and a 5.5 Earth mass planet around M dwarf stars

(Beaulieu *et al.*, 2006) provide the first indications of what may be rocky "terrestrial"

planets orbiting a main sequence star.

The techniques used for these discoveries are unlikely to detect planets of Earth's

mass around other stars. Space telescope missions will be required to take that final step,

and in the next ten to twenty years, we will not only find terrestrial planets of comparable

size to the Earth (via the transit method with NASA's Kepler mission, Borucki, et al.,

2003, 2005; Basri, et al., 2005; and the European Space Agency's Corot mission, Bordé,

et al., 2003), but also obtain spectra from them (via coronagraphic and interferometry

techniques with NASA's Space Interferometry Mission, Terrestrial Planet Finder, TPF,

Beichmann, et al., 1999; and the ESA's Darwin mission, Leger, 2000). Scientists will

need to determine how to interpret those spectra for signs of life. Although life abounds

in hidden places of the Earth, independent of sunlight, such as at hydrothermal vents

(Karl, et al., 1980; Campbell, et al., 2001) and 3- to 4-km deep metabasalt fractures (Lin,

et al., 2006), photosynthetic organisms, in particular, are the basis for nearly all life on







# DRAFT, Kiang, Astrobiology

Earth and produce some of the strongest indicators of life in abundances that can be detected astronomically.

Indicators of photosynthetic life include both distinct reflectance spectra of the organisms and gaseous products that are primarily biogenic in origin. The characteristic reflectance spectra of photosynthesizers result from the combination of pigments that preferentially absorb light in particular wavelengths and the spectral effects produced by the macrostructure of the organisms. The dominant such spectral signature on Earth is that of green plants, which absorb in the visible over about 400-700 nm, with slightly lower absorbance in the green. A stronger signature than the green "bump" is a high reflectance in the near-infrared (NIR) over about 700-850 nm, which causes a steep contrast between the visible and NIR reflectance in the red, the "red edge" (Grant, 1987; Tucker, 1976).

The notable gaseous product of photosynthesis is oxygen, the result of the splitting of water molecules (biochemistry: Rutherford and Boussac, 2004; atmospheric $O_2$ build-up: Sleep, 2001, Catling, et al., 2001; review: Wydrzynski and Satoh, 2005). The Earth's abundant atmospheric oxygen concentration is the result of the long-term balance against geochemical sinks (Kasting, 2001). Although oxygen can be produced abiotically through photolysis and the net effects of carbon burial and hydrogen escape, it is still a strong biomarker, since the significant build-up of oxygen in a planet's atmosphere and its simultaneous presence with reduced gases is unlikely without production by complex biotic processes that maintain such atmospheric chemical disequilibrium (Lovelock, 1965; Leger, et al., 1999). Other biogenic gases exist, some being direct products from organisms, such as certain volatile organic compounds emitted





# DRAFT, Kiang, Astrobiology

by plants (Guenther, et al., 1995; Kesselmeier and Staudt, 1999), and some beingsecondary products, such as $O_3$ from photolysis of $O_2$, and $NO_x$, $N_2O$, $CH_3Cl$, and COS from the breakdown of organic matter (Schlesinger, 1997; Watts, 2000). Except for $NO_x$, which has a short atmospheric lifetime and comparable abiogenic sources (Wayne, 1991), these compounds could be potential biomarkers. However, the "red edge" and oxygen remain the primary indicators of life as they result directly during the process of photosynthesis itself. Could photosynthesis arise on another planet? And if so, would it exhibit the same kind of surface spectral signature as well as the same gaseous products as on Earth?

Previous workers have examined the question of photosynthesis on extrasolar planets and its detectability. Seager, et al. (2005) reviewed the various early spectroscopic investigations that ruled out the seasonal darkening on Mars as due to vegetation, based on comparison to chlorophyll and red edge spectra. Franck, et al. (2001) and Cockell and Raven (2004) reviewed the limits of habitability for photosynthesis, from extreme environments to favorable, protective microhabitats, such as crystalline rocks that shield against UV radiation. Wolstencroft and Raven (2002) concluded that oxygenic photosynthesis on other planets is indeed plausible, even for different types of parent stars with different radiation spectra and flux densities. Heath, et al. (1999) examined in particular the distribution of temperature and radiation zones suitable for Earth-like plant photosynthesis on the surface of tidally locked planets around M dwarfs. They identified climatic regions that are favorable and summarized potential adaptations for survival or even utilization of UV flares. A number of workers have proposed that phototrophy utilizing $H_2S$ could exist in clouds on Venus (Sagan, 1961;





DRAFT, Kiang, Astrobiology

Grinspoon, 1997; Schulze-Makuch, et al., 2002; Schulze-Makuch, et al., 2004). Woolf, et al. (2002), Arnold, et al., (2002), Montañes-Rodriguez, et al. (2004, 2005), and Seager, et al. (2005), examined the Earthshine spectrum (the Moon's reflectance of the Earth's radiance) as an example of a planetary disk-averaged spectrum that might be seen by TPF. Oxygen is abundant enough in the atmosphere to be clearly observed in the Earthshine. Ozone and the oxygen dimer, $O_4$, are more difficult to detect, but present. Some of these observations show that the contrast caused by the red edge in the Earth's disk-averaged reflectance spectrum is significant enough to indicate the presence of vegetation, while Montañes-Rodriguez, et al. (2005) found the red edge obscured by clouds. A model of Earth's disk-averaged radiance spectrum by Tinetti, et al. (2006a) indicates that vegetation must cover at least 20 percent of the planet's cloud-free surface to be detectable. While the "red edge" is the main reflectance signature for photosynthesis on Earth, Wolstencroft and Raven (2002) postulated that extrasolar vegetation may have photosynthetic pigments adapted to different stellar types, possibly utilizing more photons per carbon fixed, such that we must also be prepared for the possibility of an "edge" that is not red. Seager, et al. (2005), therefore, advocated that efforts to detect photosynthesis on extrasolar planets take into account the possibility of edge-like signatures, but at other (unknown) wavelengths, in combination with biosignature gases, such as $O_2$ and $O_3$ (Lawson, et al., 2004; Leger, et al., 1993; Leger, et al., 1999). Tinetti, et al. (2006b) took Wolstencroft & Raven's (2002) suggestion that oxygenic photosynthesis could utilize three photons instead of two and posed a hypothetical extrasolar vegetation reflectance spectrum for planets around M stars by shifting an Earth vegetation reflectance spectrum so that the edge is in the NIR. They







DRAFT, Kiang, Astrobiology

found that an edge-like feature in the NIR could be easier to detect through clouds in a $CO_2$-rich atmosphere -- as they simulated for a planet orbiting an M-star -- but harder to detect in an Earth-like atmosphere.  More studies need to be done to determine whether such edge-like features are biologically reasonable in the NIR.

The above work has established the plausibility of oxygenic photosynthesis and its detectability on extrasolar planets, and acknowledges the possibility of alternative surface spectral signatures of photosynthetic organisms adapted to other radiation regimes.  Here, we will extend this work to explore in more detail the coevolution of photosynthesis—not only oxygenic but also non-oxygenic—with a different atmosphere and parent star, and we will identify how we might predict reasonable (or rule out unreasonable) surface photosynthetic spectral signatures consistent with another planetary environment. This is the next step toward discerning likely photosynthetic signatures for extrasolar planets.  With upcoming space telescope missions designed to detect and characterize planets, a better understanding of how photosynthesis could coevolve with different atmospheres and parent stars is needed to interpret the data and will help to strategize the selection of stars for detailed observation.

Our overarching tenets are these:  The light spectrum incident at a planet's surface is a function of not only the parent star's light spectrum, but also the medium through which it is transmitted—the atmospheric or water environments.  Further, the spectral transmittance of these media is a function of biogenic inputs that influence the composition of the media with further alterations by photochemistry or other biotic and chemical transformations.  Because life alters its environment, the properties of photosynthetic organisms may be partially a result of their own products or those of their







DRAFT, Kiang, Astrobiology

predecessors. Therefore, characterization of the light regime for photosynthesis must

consider spectral radiation as transmitted through a medium altered by biogenic inputs.

In this paper, we 1) formulate specific rules for the above themes for searching for

extrasolar photosynthetic surface spectra; 2) identify plausible combinations of parent

star, atmospheric composition, surface gas flux, and photosynthetic pigment properties;

3) model the surface incident photon flux densities for these scenarios; 4) identify

wavelengths for likely peak pigment absorbance and bands for photosynthetically active

radiation; and 5) quantify the potential global productivity of extrasolar photoautotrophs.

## 1.2.    Principles for photosynthetic pigment adaptation to different parent
stars and atmospheres

To pose plausible combinations of photosynthetic pigment spectra, parent star,

atmospheric composition, and surface gas flux, and to address what to look for in

extrasolar photosynthetic biosignatures, we need to consider and extend our

understanding of Earth-based photosynthesis.  In a companion paper, Kiang, et al. (2007),

we reviewed pigment and surface reflectance properties across the range of Earth's

photosynthetic organisms, from anoxygenic purple bacteria to vascular plants.  We

related these spectra to the mechanisms of light harvesting and the spectral light

environment of these organisms.  We observed how photosynthetic pigment spectra are

strongly correlated with features of the incident photon flux spectrum, such as oxygen

and water vapor bands in the atmosphere, and NIR transmittance through other organisms







DRAFT, Kiang, Astrobiology

in murky water depths.  In that review, we thus proposed rules to explain the wavelengths

at which photosynthetic pigments have their peak absorbance on Earth.

To extend those rules now for photosynthesis on extrasolar planets, we need to

accommodate the possibility that either anoxygenic or oxygenic photosynthesis might

dominate on a planet.  With regard to the latter, Wolstencroft and Raven (2002) proposed

that, on planets with little visible light, oxygenic photosynthesis could be achieved by the

utilization of more photons at longer wavelengths.  We elaborate next on how this might

be possible.

## 1.3. Potential for oxygenic photosynthesis at longer wavelengths

Several workers have observed that the potential for oxygen production, while

limited on Earth principally to photosystems that operate in the 400-700 nm range, can

theoretically extend to utilization of lower-energy photons at wavelength  as long as

1400-1500 nm if 3-4 photons are used instead of the current two in Photosystems I and II

(PS I and PS II) of oxygenic photosynthesis (Hill and Bendall, 1960; Hill and Rich, 1983;

Heath, et al, 1999; Wolstencroft and Raven, 2002).  The main requirement is that +479.1

kJ mol$^{-1}$of chemical energy be stored to allow for $CO_2 + H_2O$ --> (1/6) glucose + $O_2$.

How could the higher quantum use be achieved?

The Z-scheme of Earth-based oxygenic photosynthesis (see diagram in Kiang et

al., 2007) demonstrates how photosystems linked in series can use more than one photon

to achieve the necessary redox spans between the ground and excited states of reaction

centers to generate reduced product.  Furthermore, the oxidation of water does not

depend on the wavelengths of the photons used.  Instead, it relies on the potential





DRAFT, Kiang, Astrobiology

difference of water from the oxygen evolving complex and P680, which results from their

molecular configuration.  Since redox span, rather than reaction center midpoint redox

potential, is what constrains the photon wavelength that is useful, there is no reason

sufficient redox spans for generation of reduced product cannot theoretically be achieved

by chaining more photosystems together (i.e. a multiple Z-scheme) that utilize lower-

energy photons.  The efficiency would be influenced by the amount of illumination and

losses to favor forward over backward reactions.  It may be that a reaction center relative

to an extrasolar oxygen-evolving complex cannot achieve a sufficiently high midpoint

redox potential without some configuration that absorbs only more energetic photons.  To

date, however, we know of no theoretical constraint.  And so it would not be beyond an

extrasolar photosynthesizer, as far as we know, to evolve other kinds of electron carriers

to accommodate a higher quantum requirement.

The efficiency of conversion of light energy to chemical energy in photosynthesis

implies that there is no clear upper bound to wavelengths that could be used in

photosynthesis.  Currently, 4 photons work in concert in PS II to oxidize $2H_2O$ to evolve

an $O_2$ (Krishtalik, 1986;  Tommos and Babcock, 2000; Ferreira, et al., 2004;  McEvoy, et

al., 2005), and these are then matched with 4 photons in PS I, such that a minimum of 8

photons total are necessary for fixation of one $CO_2$.  The simultaneous action of the 4

photons on the oxygen-evolving complex is much more favorable energetically than if

they operated one at a time.  The energy input by these 8 photons is (by Planck's Law)

1387.2 kJ mol-1, if 680 nm is used in PS II (1339.0 V if 730 nm).  The efficiency of

energy input for typical plant photosynthesis is 479.1/(1339.0 to 1387.0) = ~35%, and in

typical conditions, with some unsuccessful photons and some used for cyclic





DRAFT, Kiang, Astrobiology

photophosphorylation and nitrogen assimilation (leading to up to 12 photons used; Govindjee, 1999), the efficiency is closer to 27% (Blankenship, 2002). The Calvin-Benson cycle is about 90% efficient in carbon fixation (Blankenship, 2002). The light reactions could theoretically approach nearly 100% availability of the photon energy, if illuminated with a mode-locked laser (but more typically, in Earth-like conditions and due to thermal equilibrium between ground and excited states of ensembles of reaction centers, about 2/3 of the photon energy is available for use; Parson, 1978). It is possible that an extrasolar photosynthesizer could achieve higher (or lower) efficiency.

Some efficiency limits might be set by losses from activation energies and configurational energy losses (Krishtalik, 1986, 2003). Also, at some long enough wavelength the photons could fail to provide electronically useful states and induce only vibrational energy and no electronic transitions, which would place an upper bound on eligible wavelengths. Above 1100 nm, radiation is all thermal, or at least cannot be measured using optical sensors, and this may be one possible upper limit. So far, we cannot precisely define what that limit is, as it depends on the molecular configuration. Quantification of this physical limit would be a good subject for a future study.

Therefore, a 3-photosystem series that utilizes wavelengths up to about 1040 nm could provide the same energy input as the two PS II and PS I systems in the visible. A 4-photosystem series that utilizes wavelengths up to about 1400 nm could also provide that same energy input, as could a 6-photosystem series that utilizes wavelengths up to 2100 nm. These would be equivalent to quantum requirements of, in order, 12-18, 16-24, and 24-36 photons per $CO_2$ fixed, or perhaps there might be a mix of these.





DRAFT, Kiang, Astrobiology

Similarly, one can infer photosystem sequences utilizing other electron donors with other minimum potential difference requirements for electrochemical decomposition. Thus, there is the possibility of multiple chaining of photosystems to perform oxygenic photosynthesis at longer wavelengths. Given no clear physical limit to useful wavelengths, the constraints may be primarily the spectral availability of radiation. We now extend our rules for the Earth to constrain plausible extrasolar photosynthetic biosignatures.

## 1.4.   Rules for plausible photosynthetic biosignatures

## 1.5.   Surface incident spectral photon flux densities

The light regime for photosynthesis must be characterized by the incident spectral *photon* flux density, not by the spectral *energy* flux density of solar radiation, because photosynthetic light harvesting and electron donor oxidation rely on the number of photons, not the total energy. The spectral photon flux will determine bands for "photosynthetically active radiation" (PAR). On Earth, PAR is a term generally used to refer to the 0.4-0.7 µm band used by plants, but we extend it more broadly to mean any radiation used for photosynthesis, including purple bacteria that absorb in the NIR as well as extrasolar photosynthesis.

Figure 1a reproduces Segura and co-workers (2003, 2005) spectra of radiation flux incident on the top of the atmosphere of Earth-like planets that have an average surface temperature of 288 K in orbits around observed F2V and K2V stars, the active M4.5V star AD Leo, and for modeled quiescent M stars with effective temperatures of





DRAFT, Kiang, Astrobiology

3100 K and 3650 K (M5V and M1V), respectively. In addition, the flux spectrum of the Sun incident at the top of the Earth's atmosphere is shown (sources same as in Figure 1 of Kiang, et al., 2007). The fluxes below 280 nm were not plotted because these wavelengths are not considered photosynthetically active and the surface fluxes at these short wavelengths are either negligible with an ozone shield on F, G, and K star planets, or very small on M star planets (Segura, et al., 2005, Fig.10). Figure 1a shows the energy flux density, whereas Figure 1b shows the photon flux density, which dramatizes the difference in using one spectrum versus the other to speculate on pigment adaptations. The Sun's peak energy flux density at the top of the Earth's atmosphere occurs at 450 nm, whereas its peak photon flux density occurs at 572-584 nm. In Kiang, et al. (2007), we showed how the surface incident spectral *photon* flux density could explain why chlorophyll favors the red and why plant reflectance in the green is not due to sub-optimal use of solar energy

### 1.6. Photosynthetic pigments

Photosynthetic pigments are adapted to suit several constraints: light harvesting within the available incident photon flux spectrum and oxidation-reduction potential differences to generate reduced product. In Kiang, et al. (2007), we proposed an explanation as to why chlorophyll and other pigments favor particular wavelengths, and our rationale can be applied to other planets. In particular, photosynthetic pigments would have their absorbance peaks in these wavelength categories:





DRAFT, Kiang, Astrobiology

a. the wavelength of peak incident photon flux within a radiation transmittance window, as the main environmental pressure;

b. the longest wavelength within a radiation window for core antenna or reaction center pigments, due to the resonance transfer of excitation energy and an energy funneling effect from shorter to longer wavelengths;

c. the shortest wavelengths within an atmospheric window for accessory pigments, also due to resonance transfer.

The energy per photon may limit the useful wavelength range over which sufficient redox potential differences can be achieved by the excited state of a photosynthetic reaction center; however, as we pointed out earlier, we cannot yet propose a rule for a theoretical upper limit.

### 1.7. Whole organism reflectance

Reflectance spectra of whole organisms derive from both pigment absorbance spectra and the macrostructure (e.g. cell structure, leaf morphology, canopy structure) of the organism adapted for other survival needs (e.g. climate and chemical environment). The whole organism reflectance spectrum is what ultimately can be observed by remote detection.

In our review of photosynthetic reflectance spectra (Kiang, et al., 2007), we are not able to draw conclusions about the functional role of a whole organism's reflectance





DRAFT, Kiang, Astrobiology

spectrum besides pigment influences. The variability of the red edge and the high reflectance of plants in the NIR is not well understood as an adaptation for survival, yet this is thus far the principal empirical means used to identify the presence of plants. We found the NIR reflectance not to be universally the same among all photosynthetic organisms, with a sloping "edge" for lichens, an NIR edge in purple bacteria, and trends across taxa in the wavelength at which NIR reflectance begins to plateau in the red edge, from reddest for terrestrial vascular plants, to bluer for more primitive organisms like mosses and lichens, and bluest for snow algae. More research is needed to explain these trends. A weak or sloping spectral reflectance would challenge the detectability of a reflectance signature, and given a lack of understanding about Earth-based organisms, we cannot yet conjecture whole-organism reflectance spectra for extrasolar photosynthesis.

### 1.8. Biogenic inputs influence the light spectrum

Biogenic products and, in particular, oxygen will influence the spectral transmittance of the atmosphere or water bodies. On an extrasolar planet, either oxygenic or anoxygenic photosynthesis might dominate, and oxygenic photosynthesis is possible with longer than visible wavelengths, as described earlier. Planets around cooler stars that emit little visible light could still possibly produce oxygen as a biosignature of photosynthesis. On the other hand, non-oxygenic photosynthesis was the earliest form on the early Earth and, if abundant at the global scale on another planet, could lead to a very different atmospheric composition and surface spectral signature.

We listed biogenic gases earlier: $O_2$, $NO_x$, $N_2O$, $CH_3Cl$, COS. In addition, $CH_4$ can be biogenic. These gases are not exclusively biogenic, but on Earth, biota and their





**DRAFT**, Kiang, Astrobiology

byproducts are their major sources.  The photochemistry and spectral transmittance of these gases, particularly the impact of oxygen, ozone, and UV on their abundance, should be taken into account in examining light spectra available for photosynthesis.

### 1.9.        Detectability depends on productivity

Detectability of photosynthesis depends on productivity, which depends on available resources:  light, water, electron donors, and nutrients.  We will be concerned here with the occurrence of photosynthesis on extrasolar planets that will be both abundant and distinct enough from other confounding spectra to be detectable.  Extreme environments and favorable microenvironments could support photosynthesis; however, photosynthesis at these extremes and in these hidden niches would not likely be productive enough to be detectable with a mere disk-averaged resolution of a planet, as will be available from space telescopes like TPF and Darwin.   Therefore, we focus on photosynthetic processes that can plausibly be widespread over a planet's surface.

Low global productivity by anaerobic organisms (Canfield, et al., 2006) may make these organisms difficult to detect with regard to the magnitude of their trace gas fluxes.  Their surface spectra, on the other hand, could be detectable given enough biomass build-up.  Anaerobic metabolisms may have difficulty obtaining enough energy to support multicellular organisms (Catling, et al., 2005) and, hence, to achieve abundant enough biomass to be detectable.  However, multicellular, anaerobic worms that live on a sulfur cycle exist in some aquatic environments (Woyke, et al., 2006).  This worm does not live completely independently of oxygen, as the sulfate it uses must come from aerobic respiration by some of its microbial symbionts.  Although on Earth there is no





DRAFT, Kiang, Astrobiology

multicellular photosynthetic autotroph using sulfide rather than water as the reductant, it is not clear that one could not emerge on another planet. Until we know precisely the energetic constraints to multicellular structure for autotrophs, we will assume the plausibility of complex anoxygenic plant life on land.

To summarize, given the parent star photon flux spectrum, and atmospheric and water spectral transmittance as influenced by biogenic inputs from either oxygenic or anoxygenic photosynthesis, we can identify plausible wavelengths at which extrasolar photosynthetic pigments would peak in absorbance. There is no clear physico-chemical upper limit to wavelengths useful for oxygenic photosynthesis, but the limits would most likely be ecological. We cannot predict the whole organism's spectral reflectance signature, given the lack of enough understanding about Earth organisms, but the pigment spectra can allow us to estimate how productive photosynthesis might be on an extrasolar planet. We next consider what to expect on these other planets.

## 1.10. Expected characteristics of photosynthesis on terrestrial planets around F, K, and M stars

The star types considered long-lived enough to support a habitable zone capable of evolving life are, in order of decreasing average temperature, F, G, K, and M stars, of which our Sun is a G star. By habitable zone, we mean the range of orbital distances from a star that will allow for the existence of liquid water on a planet (Kasting, *et al.*, 1993;







# DRAFT, Kiang, Astrobiology

Hart, 1978). We focus on terrestrial-type planets, since the existence of life forms on gas giants requires speculation beyond analogous examples on Earth.

Figure 1 shows that F stars emit relatively more UV radiation than the Sun, and K stars relatively slightly less UV and visible radiation. Main sequence M stars (also known as M dwarfs or red dwarfs; there also exist non-main sequence M giants) are of interest, because they are the most abundant stars in our galaxy, and they are spectrally much more different from our Sun than F and K stars. They are strongly non-blackbody emitters, with little flux of visible radiation and instead peaking in the NIR. About a quarter of M stars in their early life stage show high amounts of UV radiation and X-rays produced by flares and chromospheric activity, while quiescent stars without flares emit negligible UV radiation (Segura, et al., 2005). These flares pose potential danger to early organisms. Because of their dimness, the habitable zone of M stars is very close to the star (Kasting, et al., 1993), such that planets may become tidally locked (one side constantly facing the star; Joshi, et al., 1997). These differing radiation regimes of the F, K, and M stars lead to different atmospheric photochemistry (Segura et al, 2005) as well as a much different light spectrum for photosynthesis.

Contrasting scenarios are possible for detectable photosynthesis on extrasolar planets orbiting these stars. Since visible light from F and K stars is of comparable abundance to that on Earth, though shifted slightly in spectrum, oxygenic photosynthesis is likely to be the successful and dominant mode of photosynthesis. The dominant color of photosynthetic organisms might be different for various reasons: 1) the visible spectral flux densities are shifted, favoring peak absorbance at the wavelength of peak flux, with both downhill and uphill excitation energy transfer taking advantage of the





DRAFT, Kiang, Astrobiology

available light; 2) some other accident of evolution could lead to phycobilin-type

pigments occurring in surface photosynthesizers; 3) the color could be exactly the same,

due to the nature of energy funneling from the bluest, most energetic wavelengths to the

redder wavelengths in the reaction centers, and the only difference might be in relative

ratios of pigments at these ends.

On M stars with little visible but relatively high NIR radiation, oxygenic

photosynthesis may still be plausible, either with low productivity utilizing visible

wavelengths or photosystems that utilize longer wavelengths of light and perhaps more

photons per $O_2$ evolved.  Anoxygenic photosynthesis could possibly emerge onto land

surfaces, , since most M stars emit little UV radiation or reduce flare activity after their

early stage, and the bulk of the photosynthetically relevant radiation is in the infrared, , so

an ozone shield or screening pigments against UV would not be necessary..  Indeed,

anoxygenic photosynthesis could be the dominant form of photosynthesis, if there were

abundant non-$H_2O$ electron donors.  Photosynthetic organisms on planets around active

M stars with high UV flares (the UV can also come from the chromosphere during no

flare activity) could have adaptations to survive flare disturbances; some of these

adaptations have been observed in Earth organisms (e.g. protective pigments,

regenerative capacity, life cycles timed to avoid disturbances).  There will be little need

to protect against UV radiation from quiescent stars.





DRAFT, Kiang, Astrobiology

## 1.11.    Methods:  modeled surface incident photon fluxes, scenarios for F, G, K, and M stars, and estimation of productivity

        To conjecture about alternative photosynthetic pigment spectra on extrasolar planets, we modeled the surface radiation regime of terrestrial planets within the habitable zone around F, G, K, and M stars, given atmospheric compositions that might result from a planet dominated by anoxygenic or oxygenic photosynthesis.  Models, plausible scenarios, and an approach to compare productivity potential are described below.

## 1.12.    Models:  coupled atmospheric photochemical/radiative-convective model, and line-by-line radiative transfer model to calculate surface incident spectral photon flux density

        Segura, et al. (2003) and Segura, et al., (2005) modeled the impacts of radiation from F, K, and M stars on atmospheric photochemistry of terrestrial planets orbiting these stars.   Where they used their simulations to examine the atmospheric profiles of temperature, ozone, and other gases, and then to generate the radiance spectra of the planet as might be observed from space, we employed here the atmospheric compositions generated in their papers to look at the radiation spectrum transmitted through these atmospheres to the surface of the planet.

        Briefly, we summarize their model and parameters.  For stellar radiation emission, Segura, et al. (2003) and Segura, et al., (2005) utilized observed spectra of an F2V





DRAFT, Kiang, Astrobiology

(σ−Bootis, HD 128167), a K2V star (ε-Eridani, HD 22049), and two observed active M

stars (AD Leo and GJ643C), and modeled quiescent M stars with effective temperatures

of 3100 K and 3650 K (Figure 1a). For a baseline case, they placed an Earth-size planet

in the habitable zone and chose the semi-major axis of the planetary orbit to achieve a

planetary average surface temperature of 288 K, assuming a rotating planet. The 1-D

photochemical model (Pavlov and Kasting, 2002) and a radiative-convective model

(Pavlov, et al. 2000) werecoupled in Segura, et al. (2003, 2005) to model atmospheric

profiles of temperature, $H_2O$, $O_3$, $N_2O$, $CH_4$, CO, and $CH_3Cl$ for atmospheres with $O_2$,

$CO_2$, and $N_2$ mixing ratios like present Earth (PAL, present atmospheric level) at a

surface pressure of 1 atmosphere. The surface trace gas fluxes necessary to achieve these

mixing ratios on present Earth were calculated; these surface gas fluxes were then used

for further simulations equilibrated with prescribed $O_2$ mixing ratios. In the case of M

stars, only the 1PAL of $O_2$ atmosphere was studied. The planets around *active* M stars

have the same boundary conditions as the planets around F, G, and K stars. For those

planets around *quiescent* M stars, the boundary conditions on the photochemical model

had to be changed to avoid methane runaway (for details, see Segura et al. 2005). Given

the resulting atmospheric compositions, Segura and co-workers then used a high-

resolution radiative transfer model, SMART (Meadows and Crisp, 1996; Crisp, 1997), to

calculate the radiance spectra of these planets.

So, to examine scenarios of oxygenic and anoxygenic photosynthesis, we utilized

the atmospheric compositions that Segura and co-workers simulated. Instead of

calculating the planet's emitted radiance spectrum, we used the SMART model to

calculate the spectral photon flux densities at the surface of the planet. For underwater





# DRAFT, Kiang, Astrobiology

photosynthesis, we calculated photon fluxes at different water depths, using water

spectral transmittance from Segelstein (1981), Sogandares and Fry (1997), and Kou, et al.

(1993), as used in Kiang, et al. (2007).

### 1.13. Scenarios: F, G, K, M stars, 1 PAL $O_2$ and $10^{-5}$ PAL $O_2$.

Table 1 shows the modeled configurations that we study here, with two oxygen

scenarios and five different stars. For the oxygen concentrations, we used Segura and co-

workers' (2003, 2005) Earth-like mixing ratios (1PAL) and lowest $O_2$ mixing ratio, $10^{-5}$

PAL of $O_2$, which was the lowest stable concentration in the coupled model (with little

UV from the parent star to generate $O_3$ and, hence, OH to destroy methane, and without

another modeled methane sink, it is possible to get "runaway methane" in the model).

For the F and K stars, we looked at the 1PAL atmosphere, and for the M stars,

since $O_2$ largely absorbs in the visible and there is little visible radiation, we looked at the

1PAL $O_2$ x $10^{-5}$ atmospheres for anoxygenic scenarios. We did not examine intermediate

mixing ratios of $O_2$, since they do not lend more insight into the light spectrum for

photosynthesis. Also, the transition between the two extreme states is likely to be rapid,

such that the planet spends little time in the intermediate states (James Kasting, personal

communication).

In these scenarios, although we calculate other surface trace gas fluxes, we did not

attempt to calculate net surface fluxes of $O_2$ or estimate oxygenation times for the

planetary atmospheres, given the range of possibilities for photosynthetic efficiencies and

respiration, and the unknowns of carbon burial and crustal oxidation rates (Sleep, 2001;





DRAFT, Kiang, Astrobiology

Catling, et al., 2005). The prescribed $O_2$ mixing ratios used by Segura and co-workers were sufficient for our purposes.  Also, though full ecosystem processes – not only oxygen fluxes from photosynthesis, but also products of decomposition and other biogeochemical transformations -- have a net impact on the atmosphere, we did not vary the fluxes of other biogenic gases, since quantification of these sources is poor for the Earth.  It also turns out that, except for methane, these other gases do not have significant absorption bands at wavelengths below 1100 nm, and only methane and $N_2O$ have absorption bands that could be photosynthetically relevant between 1100-2500 nm. Methane's sources could be both biogenic and abiogenic, and variation of its fluxes could be the subject of a future study on the redox balance of extrasolar planets..  Focusing on just different oxygen scenarios provided enough variation in surface photon flux spectra for us to identify likely pigment characteristics and compare productivity potentials of different planets.

### 1.14.    Global average photosynthetic photon flux densities (PPFD) and productivity

To check the accuracy of the photosynthetic photon flux density (PPFD) calculations, we compared values for the Earth.  Zhang, et al. (2004) estimated the Earth's average annual 400-700 nm PPFD per surface area as $3.1 \times 10^{20}$ photons/$m^2$/s for clear sky conditions and $2.4 \times 10^{20}$ photons/$m^2$/s including cloud cover data (they calculated shortwave radiation with a 3D general circulation model and with satellite and ground data; we converted their shortwave radiation to visible PAR using a PAR/shortwave surface energy flux fraction from Judith Lean's data and mol/Joule using





DRAFT, Kiang, Astrobiology

485 nm as an average PAR wavelength). This is equal to a 24% reduction of surface

visible radiation by clouds. Ito and Oikawa (2004), in a model of global net primary

productivity using another cloudiness data set, estimated 2.1 x $10^{20}$ photons/m$^2$/s average

surface PPFD (352.2 J/yr over 132.3 x $10^6$ km$^2$ vegetated area) about 12.5% lower than

Zhang, et al. (2004). This may indicate a bias of clouds over vegetation. From these

spectra, we posit alternative photosynthetic pigment spectra and bands of

photosynthetically active radiation.

To estimate potential productivity, as it is beyond the scope of this paper to

address full ecosystem dynamics in detail, we focused on the productivity and pigment

spectra of photolithoautotrophic photosynthesizers, which are "primary producers" that

harvest light energy and fix carbon from $CO_2$, rather than heterotrophic photosynthesizers

that consume already reduced carbon products. The potential autotrophic productivity

depends on the actual light availability at the planet's surface, influenced by clouds. We

approximated the average noontime PPFD for the illuminated face of a cloudless planet

by the PPFD at a solar zenith angle of 60 degrees from the vertical planet's surface. To

simulate night, we used half the time-averaged PPFD with cloud cover (for non-tidally

locked planets). Since we do not know how clouds on other planets will differ from those

on Earth, we estimated the global average cloudy PPFD by reducing the time- and disk-

average PPFD values by Zhang, et al.'s (2004) amount of 24% to obtain the average

global non-cloud-free PPFD.

We extrapolated from Earth's land versus ocean light use efficiencies for net $CO_2$

fixation at the global scale. The maximum quantum yield for photosynthesis (both

oxygenic and anoxygenic) is 1/8 carbon per photons absorbed, or 0.125. Taking into





DRAFT, Kiang, Astrobiology

account possibly 12 photons total to include cyclic electron flow and nitrogen assimilation, the quantum yield is 0.083. Not all incident PAR photons get utilized for photosynthesis, however, due to environmental variability and resource limitations (water, nitrogen, soil nutrients). On Earth, gross primary productivity (GPP) — GPP is defined by biogeochemists as the gross amount of carbon fixed excluding respiration — on land ranges 90-120 Pg-C/yr (Cramer, 1999) for an ice-free land area of $132 \times 10^{12}$ km$^2$ (about 26% of the Earth's surface). Ocean productivity is approximately the same, but spread over 71% of the Earth's surface. Ocean productivity, in fact, is the main contributor to atmospheric oxygen through carbon burial. The annual surface incident PAR (400-700 nm) of $2.6 \times 10^{20}$ photons/m$^2$/s gives then an average quantum yield of just 0.006 fixed $CO_2$ per incident PAR photons for land and 0.002 for the ocean. The "net primary productivity" (NPP = gross fixed carbon minus autotrophic respiration) is about half the gross, but as respiration is very imprecisely understood, we will look just at GPP, which relates directly to photons used.

Given these efficiencies and assuming similar resource constraints (at least relatively) on other planets, the $CO_2$ fluxes from gross primary productivity can be compared for the different planets. We calculated this potential productivity but not net $CO_2$ flux from the surface, as we did not have enough information to estimate geochemical processes, such as carbon burial and crustal oxidation rates, and we did not calculate soil microbial respiration (Grace and Rayment, 2000; Raich and Schlesinger, 1992). We assumed Earth-like ice-free land cover and ocean surface area. Productivity also varies by climate zones (Joshi, et al., 1997), but our back-of-the-envelope global







estimates, like those of Wolstencroft and Raven (2002), were adequate for comparing the

model scenarios in lieu of modeling surface spatial heterogeneity.

### 1.15.    Results: spectral surface and underwater incident photon flux densities, wavelengths of peak photon flux.

Figure 2 shows the incident photon flux densities for the top-of-the-atmosphere,

the surface maximum, and the surface illuminated face average for the F, K, and M stars.

The wavelengths of peak surface photon flux are indicated, as well as major gas

absorption features.  Segura, et al. (2005) analyzed the radiance spectra for biogenic gas

signatures in the UV, the visible and up to 1.6 µm, and in the infrared over 5-20 µm.  In

general, the 1.6-3.0 µm region is not very radiant and, therefore, difficult to observe, but

for interest, our plots show the photon flux spectra up to 3 µm.  The plot for the F2V

planet in Fig. 2a indicates bands where $O_2$, $O_3$, $H_2O$, and $CO_2$ absorb, with these bands

common to all the 1 PAL scenarios.  For all cases, $H_2O$ strongly divides the available

radiation into clearly defined windows.  In the oxygenic scenarios, the oxygen A-band at

0.761 µm and B-band at 0.687 µm also possibly demarcate window boundaries, and in

Kiang, et al. (2007), we proposed that these are the boundaries for the chlorophyll

reaction center in PS II and its absorbance tail in the red edge NIR plateau.  These

transmittance windows are labeled in Fig. 2b:  U, for UV wavelengths 0.280-0.400 µm;

V, visible and Earth PAR, 0.400-0.761 µm;  W, 0.761-0.934 µm;  X, 0.934-1.135 µm;  Y,

1.135-1.350 µm;  Z, 1.470-1.840 µm;  and Q, 1.840-2.5 µm.  The cut-offs between the

W, X, Y, and Z windows are set where the photon flux reaches a minimum.  The K2V

planet's surface spectra in Fig. 2b show the same absorbance features as the F2V planet,





DRAFT, Kiang, Astrobiology

since the parent star spectra impact the atmospheric photochemistry similarly.  Primarily, the $O_3$ Chappuis band, weakly absorbing over ~0.5 to ~0.7 µm, has differing impacts on the two planets' surface spectra, based on its placement relative to the peak incident stellar radiation.  On the F2V planet, it causes the peak photon flux to be biased more strongly toward the blue, whereas on the K2V planet, it shifts the peak more toward the red, as for the Earth (Kiang, et al., 2007).

On the planets around M stars, Segura, et al. (2005) noted that the low UV flux results in high $CH_4$ concentrations, as can be seen in their IR spectra.  In addition,  in Figs. 2c, d, and e, we see $CH_4$ absorption at ~1.6-1.8 µm and at 2.3-2.4 µm, cropping the Z and Q windows for the M star planets.  Furthermore, $N_2O$ shows its signature at 3 places between 2.1 µm and 2.3 µm in the M star oxygenic scenarios (but not in the anoxygenic M star, or F2V and K2V scenarios).  If photons at these wavelengths longer than 1.1 µm are relevant to photosynthesis, then on planets around M stars, methane and nitrous oxide will alter the PAR spectrum and impact pigment properties.  $N_2O$ could be a strong biomarker at the 2.1 µm band, given that methane is ambiguous as a biomaker.

Given the above atmospheric transmittances, Figure 3 indicates the wavelengths of peak photon flux and approximate window edges at solar noon at the equator for these modeled scenarios:  the F2V planet at 1PAL, the Sun-Earth, K2V at 1 PAL, M1V at 1 PAL, AD Leo at 1 PAL, and M5V at $O_2$ x $10^{-5}$ PAL (source for the Sun-Earth same as Kiang, et al., 2007, Figure 1).  In addition, the locations of major gas absorption bands are indicated ($O_3$, $H_2O$, $CO_2$), and dashed lines are plotted for the $O_2$ A- and B-bands.

While the solar flux at the Earth's surface peaks in the red at about 685 nm, the peak for the F2V planet is in the violet at 451 nm.  The K2V planet has its peak photon





**DRAFT**, Kiang, Astrobiology

flux at 667 nm. The hottest M dwarf planet, M1V 3650K, peaks at 754 nm, within the red edge region. The M5V 3100K and AD Leo planets have their peak fluxes at about 1045 nm, in the X transmittance window where BChl b in *B. viridis* is able to absorb. The peaks would correspond to where extrasolar photosynthetic pigments would have their peak absorbance, according to our rule in Section 2.2.1 for the influence of incident spectral photon fluxes. Minor peaks are also indicated in the other transmittance windows on the figure. The wavelengths of peak photosynthetic photon flux density (PPFD) are summarized in Table 2a. Table 2a also summarizes the integrated PAR-relevant fluxes in different transmittance windows, which we will discuss in more detail later.

Table 2b gives the time-averaged global PPFD. The average G2V surface PPFD in Table 2b is 3.2 x $10^{20}$ photons/m$^2$/s for clear sky conditions, compared to 3.1 x $10^{20}$ in Zhang, et al. (2004). So, our estimates for the G2V planet appear fairly accurate (Wolstencroft and Raven, 2002, estimated an absorbed 14.0 x $10^{20}$ photons/m$^2$/s for a cloudless G2V planet at the inner edge of the habitable zone).

Since photosynthesis on Earth developed first under water, and it is sedimentation of fixed carbon in the ocean that primarily drives the build-up of atmospheric oxygen, we are also interested in photon flux densities under water for when the atmosphere is low in oxygen. Figure 4a shows the maximum (at solar noon at the equator) spectral flux density at a depth of 5 cm of pure water for all the star types, with the F, K, and quiescent M1V and M5V planet atmospheres at negligible $O_2$ concentrations. This illustrates how water absorbance reduces the available NIR photons on all the planets. Figure 4b shows the maximum underwater spectral flux density at several depths for the M5V planet with





DRAFT, Kiang, Astrobiology

no oxygenic photosynthesis. The lower photon flux limit is set at the level of the Black Sea and hydrothermal vent bacteria at $1.8 \times 10^{15}$ photons/m$^2$/s (this is just an indicator, since the available photon flux would be integrated over a band). Underwater photon fluxes are strongly curtailed at wavelengths above 1.34 µm for a depth of 5 cm, or above 1.14 µm at depths 20 cm and deeper.

## 1.16.   Discussion

### 1.17.        Available PAR bands and pigment spectra

Several previous workers have addressed the various star type light limits for oxygenic photosynthesis (Cockell and Raven, 2004; Wolstencroft and Raven, 2002; Franck, et al., 2001; Heath, et al. 1999), so our numbers here provide more refined calculations for comparison of photosynthetic photon flux densities (PPFD). Wolstencroft and Raven (2002) estimated surface incident PPFD and PPFD at a 10 m ocean depth for planets around A0V, F0V, G0V, G2V, K0V, and M0V stars. Our modeled stars—F2V, G2V, M1V, M4.5V, and M5V—except for the M4.5V and M5 stars, are comparable to theirs. Wolstencroft and Raven (2002) assume spectral peaks according to blackbody radiation, so here we provide the non-blackbody spectra for M stars as in of Segura, et al. (2003, 2005). With the SMART radiative transfer model (Meadows and Crisp, 1996; Crisp, 1997), we can now define spectral photon fluxes in line-by-line detail.

On these different stars, a few different radiation bands could be potentially photosynthetically active. For the F, G, and K stars, since there is abundant visible light,





DRAFT, Kiang, Astrobiology

it is most likely that a visible 400-700 nm band would be the dominant PAR band.  In contrast, for the M star planets, there is significantly higher PPFD in the NIR than in the visible compared to the planets around F, G, and K stars, with the available radiation occurring in distinct windows.  So, it could be advantageous on such planets for PAR to extend to the NIR.  If pigments evolve at the surface, then the PAR range could extend to 1800 nm and possibly to 2500 nm.

In the underwater spectra in Figure 4, the 1.14 µm upper wavelength cutoff to the X band and the 1.34 µm cutoff for only very shallow organisms above 5 cm indicate it is unlikely that photosynthesis developing under water will evolve pigments that absorb above about 1.340 µm, and most likely the pigments will absorb somewhere below 1.140 µm, the cutoff getting shorter with depth.  All aquatic and marine photosynthesizers would be restricted to below this wavelength.  Their land plant offspring may be confined to these bands, but it is also possible that pigments can evolve for new environments to absorb at wavelengths different from their ancestors.

So, for M star planets, PAR bands in the NIR could have the following restrictions:  a 1040 nm 3-photosystem upper wavelength limit for the energy equivalent required by PS II and PS I, a 1100 nm upper wavelength thermal limit, and transmittance windows where there is the most available radiation.  We can allow for the possibility that an extrasolar photosynthesizer could achieve slightly better efficiency and fix carbon with a 3-photosystem arrangement out to 1100 nm.  We, therefore, define extended PAR bands for M star planets.  These bands correspond closely to the W, X, Y, Z, and Q windows defined in Fig. 2b, rounded to the nearest tenth of a micron; since the regions at the window cut-offs in the NIR are noisy and spread out, a higher precision makes little





DRAFT, Kiang, Astrobiology

difference in the integrated photon flux. Also, since organisms absorbing at longer wavelengths will most likely also harvest light at the shorter wavelengths via resonance transfer, each PAR band includes the shorter-wavelength bands. The extended PAR bands are: 0.4-1.1 μm, 0.4-1.4 μm, 0.4-1.8 μm, and 0.400-2.5 nm. These correspond to photosystems that utilize, respectively, 1.5 times, 2 times, 3 times, and 4 times as many photons per carbon fixed as Earth organisms.

For the F, G, and K star planets, the PAR band would most likely be just the same as on Earth, 0.4-0.7 μm, with limited photosynthesis at wavelengths up to 1.1 μm. The bounds of the shorter PAR band could be set at 0.400-0.730 μm to acknowledge the ability of *Acaryochloris marina* to utilize the longer wavelengths, but for the sake of intercomparison with common numbers in the literature, we stick to the traditional 0.7 μm cutoff.

For all of the simulated stars/atmosphere combinations, the solar noon surface flux densities in all bands in Table 2a are bright enough to sustain photosynthesis above the light compensation points for the Black Sea and hydrothermal vent bacteria at 1.8 x $10^{15}$ photons/m$^2$/s (Overmann, et al., 1992; Beatty, et al., 2005), and for green plants at 1.8 x $10^{18}$ photons/m$^2$/s (Nobel, 1999). The M star planets' photon fluxes in the 0.4-1.1 μm band are more than twice the Earth's visible PPFD. Therefore, even planets around M stars could ostensibly host land-based plants. Meanwhile, the K stars and quiescent M stars produce low UV well below the range experienced by Earth organisms, even without an ozone shield.

The presence of 1 PAL oxygen on the M5V 3100K planet does not significantly lower the integrated PPFD, but it causes very narrow band dips in PPFD at 0.687, 0.761,





DRAFT, Kiang, Astrobiology

and 1.269 µm. Hence, though oxygen may demarcate PAR bands for photosynthetic pigments and slightly shift the visible spectrum, it does not greatly affect light availability compared to a planet with anoxygenic photosynthesis. The oxygen spectral influence would be more important for an M1V 3650 K planet in the visible, whereas for the cooler M stars with little visible light, it would be more important at the 1.269 µm line.

So, spectrally we have some sense for the wavelengths at which reaction centers and pigments may peak in photon absorbance. The peak PPFD wavelengths summarized in Table 2a are not necessarily at the edges of their respective transmittance windows, so possibly there may be uphill and downhill excitation energy transfer, or the RC peaks may be shifted to the edges of the conjectured PAR bands, with accessory pigments harvesting in the rest of the band. The peak at 0.754 µm for the M1V star is in a region of the spectrum that is so variable that photons in longer NIR wavelengths may still be more reliable energy sources. In the NIR, the wavelength of peak PPFD at a 5 cm depth for the M5V planet is as for Earth, at water's peak transmittance at ~1.075 µm. Since the purple bacteria that are known to absorb at 1.020 µm (which is neither a peak flux nor a window edge wavelength) use that energy for uphill transfer to a reaction center at 0.960 µm, this could be the case on an extrasolar planet as well, and more work on the thermodynamics and kinetics of exciton transfer needs to be done to determine why 0.960 µm for the RC and 1.020 µm for the accessory pigment or some other wavelength could be an optimum. The wavelength limit may have to do with the energy cut-off between electronic transitions versus vibrational or thermal energy, as mentioned earlier.





DRAFT, Kiang, Astrobiology

### 1.18.    **Productivity potential for multiple photosystems**

Having determined that our simulated planets meet minimum PPFD to support not only the most light limited organisms found on Earth but also vascular plants at typical flux levels, we now look at the productivity potential at the global scale using the values in Table 2b, adjusting to average over time and account for clouds, as described earlier.  The detectability of photosynthetic atmospheric signatures – oxygen and seasonal cycles of $CO_2$ and/or $CH_4$ -- will depend on the net global productivity of both ocean and land organisms.  The detectability of surface spectra will depend on the photosynthesizers' cover and density over the planet. As the purpose of this study is to examine planets in the habitable zone (where liquid water exists), and resource constraints on another planet may be arbitrary, we can assume sufficient water and nutrients as a plausible case, and the extreme limits of photosynthesis are of concern primarily with respect to how much they will affect the spectral properties, productivity, and hence detectability of photosynthesis.

Table 3 summarizes the global time-averaged PPFDs and the $O_2$ or $CO_2$ flux from net primary productivity, weighted for 26% ice-free land and 71% ocean for the  planets around the other stars.  The GPP values are given assuming the same quantum efficiency as on Earth, as well as using 1.5x, 2x, 3x, and 4x quantum requirements for the PAR bands that extend to the NIR.  Also, since underwater photosynthesis will have more limited useful wavelengths, only the PAR 0.4-1.1 µm and 0.4-1.4 µm bands are used for the ocean.  As a reference, Earth's GPP is given in bold (0.72  x $10^{18}$ quanta-$CO_2$/m$^2$/s is equivalent to ~110 Pg-C/yr, where quanta are molecules).





# DRAFT, Kiang, Astrobiology

In the 0.4-1.1 µm PAR band, all of the stars cooler than the Sun have photon fluxes at the planetary surface well above the visible photon fluxes at the Earth's surface. If 3-photosystem photosynthesis is used (1.5x quantum requirement over 400-1100 nm), then the M1V planet could still have a productivity above that of Earth's, but the other M stars would be only about 75 percent as productive as the Earth, at Earth-like efficiencies and the given land/ocean coverage. If a 4-photosystem series is used (2x quantum requirement over 0.4-1.4 µm), the M1V planet can still maintain a productivity close to that of Earth's, while the M4.5V and M5V planets increase slightly to 80 percent of Earth's productivity. If triple or quadruple quantum requirements are necessary, then productivity on the M4.5V and M5V planets would drop to two thirds to a half that of Earth. Multiple photosystems with graduated quantum requirements with each extension of the PAR bands to the NIR could boost the productivity, but the M4.5V and M5V planets would still be below Earth's productivity if utilizing only the 0.4-1.4 µm band; the longer bands would actually almost provide the necessary quantum flux, but also require that pigments evolve outside the water. Productivity will be limited by the availability of S, Fe, and $H_2$ (estimated for the early anoxic Earth by Canfield, et al., 2006). On planets like Venus and Mars, sulfur is quite abundant (Schulze-Makuch, et al., 2004; Wang, et al., 2006) and may not be too limiting, but its utilization outside of marine or aquatic environments would require the ingenuity of an extrasolar photosynthesizer.

How the lower productivities might affect the atmospheric signature of oxygenic photosynthesis depends on the redox chemistry and history of the planet for the buildup of oxygen. Catling, et al. (2005) reviewed the parameters that affect the oxygenation





DRAFT, Kiang, Astrobiology

time, these parameters being: planetary size, which affects tectonic activity, hydrogen

escape; the presence of continents, which affects weathering, oceanic heat flux, carbon

burial; planet composition, which determines whether conditions are more oxidizing or

reducing; and the community composition of the biosphere, which affects cycling and

burial of organic and inorganic carbon. Although we simulate planets with the same size

as and similar continental fraction as the Earth, we cannot simulate full biogeochemical

cycles to predict the build-up of atmospheric $O_2$, but we make comparisons, assuming

otherwise Earth-like ratios of sources and sinks of $O_2$. On Earth, about 0.1 Pg-C/yr is

buried in ocean sediments, or about 0.2% of ocean NPP. In Segura et al.'s (2003)

simulations of the atmospheres of planets around F2V, G2V, and K2V stars, $O_2$ is

observable in the visible for atmospheres with at least $10^{-2}$ PAL $O_2$, and $O_3$ can be

observed in the thermal-IR in atmospheres with at least $10^{-3}$ PAL $O_2$. So, if the level of

productivity is linearly related to the concentration of atmospheric $O_2$, then the dim M5V

planets could still show a strong enough biosignature.

Other scenarios of land/ocean configurations can be conceived, with land

concentrated at the equator versus the poles, variations in efficiencies of light use, etc.

The oceans and land compete, however, for the strength of biosignatures. Increasing the

size of the oceans could enhance the $O_2$ signature but then weaken the land surface

signature. Although algal blooms can create a strong signature for satellite detection

(Carder, et al., 1985; Lubin, et al., 2001), their extent at the global scale is too sparse for

detection in a disk-average spectrum (Tinetti, et al., 2006a). Meanwhile, the location of

land toward the poles or equator, or between hemispheres as well as land productivity

would determine the strength of a seasonal signal, if any. It may be difficult to discern a





# DRAFT, Kiang, Astrobiology

seasonal-type atmospheric biosignature for M4.5V to M5V planets that utilize only visible PAR for photosynthesis.

## 1.19. UV flares and photosynthesis under water

Because UV flares during the early life of M stars may push organisms to deeper water for protection, there may be a conflict between UV screening versus availability of photosynthetically active radiation (Raven and Wolstencroft, 2002, call this "feast or famine"). Here we estimate the depths of water at which organisms would be safe from flares of different magnitudes, assuming that they have no UV screening pigments. UV damage to plants has been observed at doses of 15-16 kJ/day (Kakani, et al., 2003). The spectral quality of M star flares is not well known, but they may span the X-ray, UV, and visible. Segura, et al. (2005) summarized their magnitudes and duration, and some flares from AD Leo. Flares may be as small as $\sim 10^{28}$ erg of radiative energy and occur fairly frequently ($\sim 0.71$/hr) (Crespo-Chacon, et. al., 2006); the largest observed are $10^{34}$-$10^{37}$ erg, and the flares larger than $\sim 10^{32}$ ergs occur at a rate of roughly one per day on very active stars. On AD Leo, flares of order $10^{32}$ ergs occur approximately every 4 days, and a flare of size $7.1 \times 10^{32}$ ergs was observed to have 70% of its energy in the UV in 0.200-0.326 $\mu$m (Hawley and Pettersen, 1991).

To estimate the strongest dose that organisms could receive at the surface of a planet, we looked at the doses for a planet around AD Leo, whose extremely high activity represents a possible upper limit to UV emissions (Segura, et al., 1993). Although damage from UV is more precisely estimated with action spectra for DNA damage, since these action spectra as well as flare spectra are so variable, we used integrated estimates





DRAFT, Kiang, Astrobiology

the 15 kJ/day plant dose and the average water absorbance of radiation over 0.200-0.326 µm from the spectrum of Segelstein (1981) (weighted by photon energy, this gives an absorbance of ~ 0.01 cm$^{-1}$). Given the above occurrence rates of the flares, and the semi-major axis of 0.16 AU from AD Leo, we solved for the safe depths from UV damage, at which the available UV from the flare is less than the minimum plant dose. Table 5 summarizes these depths, as well as the visible (0.4-0.7 µm) and NIR (0.7-1.1 µm) PAR photon flux density averages for the illuminated face of a rotating planet.

The smaller flares of $10^{28}$-$10^{31}$ ergs, even at their higher rate of occurrence, do not produce enough UV to be damaging even outside of the water. Also, the flares of order $10^{32}$ ergs, even at a higher rate than on AD Leo, produce below the plant damage limit at the surface. The flares of ~ $10^{33}$ ergs are at the threshold at which protection under water is necessary. The largest flares of $10^{37}$ ergs result in dangerous UV doses down to at least a 9.1 m water depth. However, even at this depth, AD Leo provides visible radiation an order of magnitude above the lower limit of 1.8 x $10^{18}$ photons m$^{-2}$ s$^{-1}$ for green plants (Nobel, 1999), and well above the red algae limit of 6 x $10^{15}$ photons m$^{-2}$ s$^{-1}$ (Overmann, et al., 1992). Although the NIR drops off significantly with depth, its photon flux at the 9.1 m water depth is still above the red algae limit; also, organisms at this depth would harvest the visible as well as NIR photons. Therefore, photosynthetic organisms during the flaring stage of M star planets should be able to survive even with visible light, though their productivity would be limited to less than 14% of Earth's with smaller flares and less than 4% with very active M stars like AD Leo with daily large flares.





DRAFT, Kiang, Astrobiology

## 1.20. Conclusions

From Earth's example, we have proposed how photosynthesis is likely to evolve with a different parent star and atmospheric chemistry, and produce predictable alternative pigment spectra and atmospheric signatures. We do not know enough yet to predict full reflectance spectra.

During our survey of Earth-based photosynthetic organisms, we observed a number of molecular-scale chemical constraints and global scale environmental pressures determine the optimum wavelength of peak absorbance of photosynthetically active radiation (Kiang, et al. 2007). Photosynthetic pigments on extrasolar planets are likely to evolve to peak in absorbance at the wavelengths at which the transmitted light at the planet's surface is highest in photon flux densities or most energetic for transfer to longer-wavelength reaction centers. Oxygenic photosynthesis could be possible over wavelengths up to even beyond 2.5 µm, given a sufficient number of photosystems chained together and barring a longer wavelength limit for vibrational versus electronic energy transitions. However, pigments that evolve under water are unlikely to absorb above 1.4 nm or even 1.1 nm. Anoxygenic photosynthesis would have the same PAR band restrictions. Planets around F2V stars might have a "blue edge" or at least be darker in the blue. K2V planets would look very similar to the Earth. The M stars could have several different PAR bands: they might have multiple critical absorption wavelengths in the visible, due to their noisy spectra in this region; the dominant photosynthetic organisms would most likely harvest light over 0.4-1.1 µm, with potential but unlikely extensions to 1.4 and 2.5 µm.





DRAFT, Kiang, Astrobiology

All of the star-planet combinations we simulated have more than sufficient light to support Earth-like photosynthesis. The low productivity of the M4.5V and M5V planets, if oxygenic, could be sufficient to produce observable oxygen, but probably take longer to do so. Redox energetics imply that longer wavelength photons could be used in chains of several linked photosystems to provide enough energy to abstract electrons and fix $CO_2$.

For the purpose of observing terrestrial extrasolar planets, we cannot yet determine how strong the red edge contrast will likely be for detection purposes, and given the example of lichens, there might not be a distinct edge. Remote sensing studies indicate that lichens can be distinguished from the mineral background (Ager and Milton, 1987; Rees, et al., 2004), so global scale modeling like that of Tinetti, et al. (2006a and 2006b) will help us to determine how detectable a lichen-like signature might be.

This investigation of the light spectrum at the surface of planets around other star types offers the next refinement in quantifying the nature of photosynthesis on extrasolar planets, particularly around M stars. Here we focused on predicting and identifying photosynthetic biosignatures through narrowing the possibilities for pigment absorbance spectra. We have identified where deeper understanding of photosystem energetics is needed to restrict potential pigment absorbance wavelengths. Because more data are needed to explain the apparent variation in the NIR reflectance spectra across organism functional types, particularly algae and bacteria, we have *not* attempted to simulate a full surface reflectance spectrum and its detectability in the radiance spectrum of the planet. More studies are needed to assess the likelihood of, for example, the edge-like reflectance





# DRAFT, Kiang, Astrobiology

feature of the hypothetical NIR-shifted reflectance spectrum that Tinetti, et al. (2006b)

simulated for photosynthesis around M stars.

Additional modeling investigations we feel are called for include: 3D modeling

to refine calculations of photosynthetic productivity by climate zones and as affected by

cloud cover; more 3D radiative transfer modeling to detect alternative pigments and

whole-canopy reflectance spectra; coupled biogeochemical/atmosphere models to

quantify net fluxes to the atmosphere and the possible amplitude in temporal variation.

For instrument design, modeling is needed to determine the spectral resolution necessary

to resolve a pigment feature in a time- and disk-averaged radiance spectrum of the planet

(e.g. Tinetti, et al., 2006b, examine the red edge detectability on Earth). Future work is

also needed to assess an optimum resolution to capture pigment spectra that could be as

sharp as that of terrestrial plants or as gradual as those of lichens or purple bacteria. If

photosynthesis occurs in sulfur-rich clouds as on Venus, then even more complex

modeling of cloud spectral radiative transfer would be another challenge. Finally, a

survey should be conducted of mineral types to distinguish extrasolar photosynthetic

signatures from the mineral surface.

Biosignatures -- both atmospheric and surface -- on planets around M stars may

actually be easier to detect than those around F, G, or K stars. The modeled atmospheres

of M star planets of Segura, et al. (2005) reveal that low UV radiation from quiescent M

stars could result in higher concentrations of biogenic gases $CH_4$, $N_2O$, and $CH_3Cl$.

Tinetti, et al. (2006) found the "red edge," shifted to the NIR, to be easier to detect

through modeled clouds than the plant red edge. These prospects to detect life should





DRAFT, Kiang, Astrobiology

motivate continued investigations into M star atmospheres and the spectral adaptations of extrasolar photosynthesis.

## Acknowledgments

We are greatly obliged to Niels-Ulrik Frigaard, Richard Cogdell, and Andrew Gall for pigment data and valuable comments; Brian Cairns, Judith Lean, and Andrew Lacis for solar spectral photon flux densities for the Earth; David Mauzerall, Warwick Hillier, Yongqin Jiao, and Elmars Krausz for very helpful explanations about photochemistry. We also thank John Scalo, Norm Sleep, and Jim Kasting for many lively discussions and helpful references. Thanks are due to two anonymous reviewers who helped greatly to improve this paper. This work was performed as part of the NASA Astrobiology Institute's Virtual Planetary Laboratory Lead Team effort, supported under NASA Cooperative Agreement No. CAN-00-OSS-01. M.C. thanks NASA for supporting his participation in this work through JPL contract 1234394 with UC Berkeley. Govindjee thanks the Dept. of Plant Biology of the University of Illinois for office support. N.Y.K also thanks NASA and James Hansen for supporting this work.

# Table captions

Table. 1.  Model scenarios:  prescribed conditions, oxygen levels, parent stars.

Table 2a.  Maximum surface and underwater incident photon flux densities and wavelengths of peak photon flux density at solar noon at the equator, for cloudless planets.

Table 2b.  Average surface and underwater incident photon flux densities for the illuminated face of cloudless planets.  Average fluxes are approximated by the photon flux at a solar zenith angle of 60 degrees from vertical.

Table 3.  Global time-averaged surface and underwater incident photon flux densities and potential gross primary productivity (GPP) for planets with oxygenic or anoxygenic photosynthesis, and for cases of photosystem quantum requirement that are 1) equal to that on Earth, and 2) with double, triple, and quadruple the quantum requirements, respectively, for the 400-1100 nm, 400-1400 nm, 400-1800 nm, and 400-2500 nm PAR bands.

Table 4.  Water depths safe from UV flares;  available visible (0.4-0.7 microns) and NIR (0.7-1.1 microns) for AD Leo.





**Figure Captions**

Figure 1.  Top-of-the-atmosphere incident radiation for F2V, G2V (Sun), K2V, M1V, M4.5V, and M5 stars, in terms of a) energy units (reproduced from Segura, et al., 2005), and b) photon flux densities.

Figure 2.  Incident spectral photon flux densities at the top of the atmosphere, at the surface at the equator at solar noon, and for the surface illuminated face, clear-sky average for an Earth-like planet with Earth's present atmospheric level (1 PAL) of oxygen and other biogenic gas surface fluxes, as well as with atmospheric oxygen at $10^{-5}$ x 1 PAL.  Planetary atmospheres calculated by Segura, et al. (2003, 2005) for the following parent star types:  a) F2V, b) K2V, c) M1V, d) M5V, and e) AD Leo, an active M4.5V star.  Wavelengths of peak photon flux and gases with significant absorption bands are indicated.

Figure 3.  Surface incident photon flux densities for the Earth, and for planets in the habitable zone of F, K, and M stars (temperatures of 3650 K, 3100 K, and measured AD Leo), as calculated from atmospheric composition in Segura, et al. (2003), Segura, et al. (2005), and the SMART radiative transfer model (Crisp, 1997).  Wavelength of peak flux densities and transmittance window edges are indicated, as well as $O_2$ absorption lines. The absorbance spectra of BChl a and BChl b are included.

Figure 4.  Underwater planetary maximum photon flux densities for star-atmosphere scenarios in Table 2.  a) Photon fluxes at 5 cm depth in pure water.  b) Photon fluxes at





the surface and several depths for an M5V 3100K star with $O_2$x$10^{-5}$ PAL.







Table 1.  Model scenarios:  prescribed conditions, oxygen levels, parent stars.
(PAL=present atmospheric level of Earth.)

| Prescribed conditions | Photosynthesis scenarios | Parent Stars |
|---|---|---|
| * 288 K Earth-size planet surface temperature<br>* Other gases: $CO_2$, $CH_4$, $N_2O$, CO calculated surface fluxes for present Earth mixing ratios (1PAL) | 1. Oxygenic photosynthesis $O_2$ = 1 PAL mixing ratio<br><br>2. Anoxygenic photosynthesis $O_2$ = 1 PAL $O_2$ x $10^{-5}$ | 1. F2V modeled<br>2. K2V modeled<br>3. M4.5V known active: AD Leo<br>4. M5V modeled quiescent : T=3100K<br>5. M1V modeled quiescent: T=3650K |





Table 2a. Maximum surface and underwater incident photon flux densities for cloudless planets at solar noon.

| | F2V 1PAL | G2V Sun/Earth | K2V 1PAL | M1V 1PAL | M4.5V 1PAL | M5V O₂x1e-05 | M5V 1PAL | M5V O₂x1e-05 5 cm | M5V O₂x1e-05 100 cm |
|---|---|---|---|---|---|---|---|---|---|
| | | | | | | | (x10²⁰ photons/m²/s) | | |
| UV-B 280-315 nm | 0.049 | 0.018 | 0.015 | 0.001 | 0.000 | 0.000 | 0.000 | - | - |
| UV-A 315-400 nm | 2.319 | 0.871 | 0.588 | 0.095 | 0.021 | 0.016 | 0.016 | - | - |
| PAR 400-700 nm | 16.4 | 11.0 | 11.5 | 6.1 | 1.8 | 1.5 | 1.5 | 1.4 | 1.1 |
| PAR 400-1100 nm | 29.8 | 23.8 | 26.3 | 23.2 | 16.1 | 17.3 | 16.9 | 9.9 | 1.5 |
| PAR 400-1400 nm | 34.0 | 28.6 | 32.1 | 29.3 | 23.7 | 25.7 | 24.9 | 10.2 | 1.5 |
| PAR 400-1800 nm | 37.9 | 33.7 | 38.9 | 36.4 | 34.2 | 35.3 | 34.3 | 10.2 | 1.5 |
| PAR 400-2500 nm | 40.5 | 36.9 | 43.3 | 40.2 | 40.0 | 40.1 | 38.4 | 10.2 | 1.5 |
| peak photon flux wavelength range (nm) | 450.8 451.0 | 668.5 685.5 | 666.6 667.8 | 754.3 753.5 | 1045.1 1045.9 | 1042.8 1043.6 | 1042.8 1043.6 | 1073.2 1075.7 | 1073.2 1075.7 |





Table 2b. Average surface and underwater incident photon flux densities for cloudless planets, illuminated face.

| | F2V 1PAL | G2V Sun/Earth | K2V 1PAL | M1V 1PAL | M4.5V 1PAL | M5V $O_2$x1e-05 | M5V 1PAL | M5V $O_2$x1e-05 5 cm | M5V $O_2$x1e-05 100 cm |
|---|---|---|---|---|---|---|---|---|---|
| | | | | | (x$10^{20}$ photons/m$^2$/s) | | | | |
| UV-B 280-315 nm | 0.006 | 0.017 | 0.003 | 0.000 | 0.000 | 0.000 | 0.000 | - | - |
| UV-A 315-400 nm | 0.940 | 0.518 | 0.241 | 0.040 | 0.009 | 0.007 | 0.007 | - | - |
| PAR 400-700 nm | 7.5 | 6.3 | 5.4 | 2.9 | 0.9 | 0.7 | 0.7 | 0.7 | 0.5 |
| PAR 400-1100 nm | 13.8 | 12.4 | 12.3 | 10.8 | 7.4 | 7.8 | 7.8 | 4.7 | 0.7 |
| PAR 400-1400 nm | 15.6 | 14.2 | 14.8 | 13.2 | 10.6 | 11.0 | 11.0 | 4.8 | 0.7 |
| PAR 400-1800 nm | 17.4 | 16.4 | 17.9 | 16.2 | 15.0 | 15.0 | 15.0 | 4.8 | 0.7 |
| PAR 400-2500 nm | 18.5 | 17.6 | 19.8 | 17.7 | 17.3 | 16.5 | 16.5 | 4.8 | 0.7 |





Table 3. Time- and disk-averaged PAR, GPP.

| | F2V 1PAL | G2V Sun/Earth | K2V 1PAL | M1V 1PAL | M4.5V 1PAL | M5V $O_2$x1e-05 | M5V 1PAL |
|---|---|---|---|---|---|---|---|
| Time- and disk-averaged surface incident radiation (x$10^{20}$ photons/m$^2$/s) | | | | | | | |
| UV-B 280-315 nm | 0.002 | 0.006 | 0.001 | 0.000 | 0.000 | 0.000 | 0.000 |
| UV-A 315-400 nm | 0.357 | 0.197 | 0.092 | 0.015 | 0.003 | 0.003 | 0.003 |
| PAR 400-700 nm | 2.8 | 2.4 | 2.0 | 1.1 | 0.3 | 0.3 | 0.3 |
| PAR 400-1100 nm | 5.2 | 4.7 | 4.7 | 4.1 | 2.8 | 3.0 | 3.0 |
| PAR 400-1400 nm | 5.9 | 5.4 | 5.6 | 5.0 | 4.0 | 4.2 | 4.2 |
| PAR 400-1800 nm | 6.6 | 6.2 | 6.8 | 6.2 | 5.7 | 5.7 | 5.7 |
| PAR 400-2500 nm | 7.0 | 6.7 | 7.5 | 6.7 | 6.6 | 6.3 | 6.3 |
| GPP with Earth-like quantum requirements (x$10^{18}$ quanta/m$^2$/s) | | | | | | | |
| PAR 400-700 nm | 0.85 | **0.72** | 0.61 | 0.33 | 0.10 | 0.08 | 0.08 |
| PAR 400-1100 nm | 1.56 | 1.40 | 1.39 | 1.22 | 0.84 | 0.88 | 0.88 |
| PAR 400-1400 nm | 1.76 | 1.60 | 1.68 | 1.50 | 1.20 | 1.25 | 1.25 |
| PAR 400-1800 nm | 1.87 | 1.74 | 1.86 | 1.68 | 1.46 | 1.48 | 1.48 |
| PAR 400-2500 nm | 1.94 | 1.81 | 1.97 | 1.76 | 1.59 | 1.57 | 1.57 |
| GPP at 1.5x, 2x, 3x, and 4x quantum requirements (x$10^{18}$ quanta/m$^2$/s) | | | | | | | |
| PAR 400-1100 nm | 1.04 | - | 0.93 | 0.81 | 0.56 | 0.59 | 0.59 |
| PAR 400-1400 nm | 0.88 | - | 0.84 | 0.75 | 0.60 | 0.62 | 0.62 |
| PAR 400-1800 nm | 0.62 | - | 0.62 | 0.56 | 0.49 | 0.49 | 0.49 |
| PAR 400-2500 nm | 0.48 | - | 0.49 | 0.44 | 0.40 | 0.39 | 0.39 |





Table 4. Water depths safe from UV flares;  available visible (0.4-0.7 microns) and NIR (0.7-1.1 microns) for AD Leo.

| flare freq | flare size | safe depth | (photons/m$^2$/s x 10$^{20}$) M4.5V AD Leo | |
|---|---|---|---|---|
| (d$^{-1}$) | (ergs) | (m) | VIS | NIR |
| 17 | 1.0E+28 | 0.0 | 0.9 | 6.6 |
| 17 | 1.0E+29 | 0.0 | 0.9 | 6.6 |
| 17 | 1.0E+30 | 0.0 | 0.9 | 6.6 |
| 17 | 1.0E+31 | 0.0 | 0.9 | 6.6 |
| 1 | 1.0E+32 | 0.0 | 0.9 | 6.6 |
| 1 | 1.0E+33 | 0.0 | 0.9 | 6.6 |
| 1 | 1.1E+33 | 0.05 | 0.8 | 3.7 |
| 1 | 5.0E+33 | 1.5 | 0.6 | 0.1 |
| 1 | 1.0E+34 | 2.2 | 0.5 | 3.E-02 |
| 1 | 1.0E+35 | 4.5 | 0.3 | 3.E-03 |
| 1 | 1.0E+36 | 6.8 | 0.3 | 6.E-04 |
| 1 | 1.0E+37 | 9.1 | 0.2 | 1.E-04 |





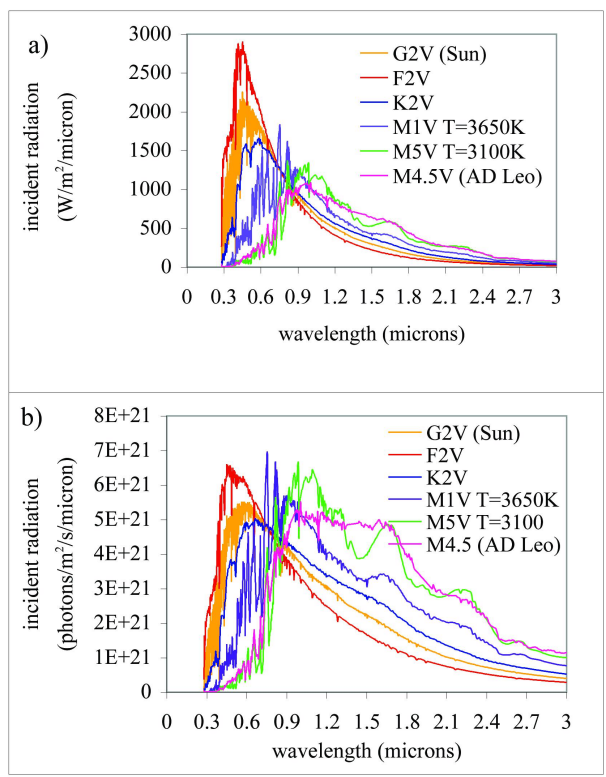

Figure 1ab. Top of the atmosphere incident radiation.

**Figure 1 Top of the atmosphere incident radiation.**





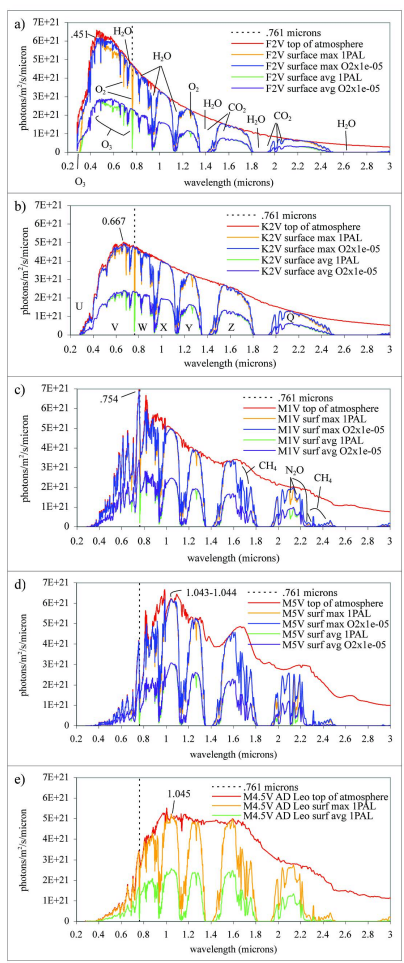

Figure 2

**Figure 2 Incident spectral photon flux densities at the top of the atmosphere, at the surface at the equator at solar noon, and for the surface illuminated face, clear-sky average for an Earth-like planet with Earth's present atmospheric level (1 PAL) of oxygen and other biogenic gas surface fluxes, as well as with atmospheric oxygen at $10^{-5}$ x 1 PAL. Planetary atmospheres calculated by Segura, et al. (2003, 2005) for the following parent star types: a) F2V, b) K2V, c) M1V, d) M5V, and e) AD Leo, an active M4.5V star. Wavelengths of peak photon flux and gases with significant absorption bands are indicated.**





Figure 3.

**Figure 3 Surface incident photon flux densities for the Earth, and for planets in the habitable zone of F, K, and M stars (temperatures of 3650 K, 3100 K, and measured AD Leo), as calculated from atmospheric composition in Segura, et al. (2003), Segura, et al. (2005), and the SMART radiative transfer model (Crisp, 1997). Wavelength of peak flux densities and transmittance window edges are indicated, as well as O2 absorption lines. The absorbance spectra of BChl a and BChl b are included.**





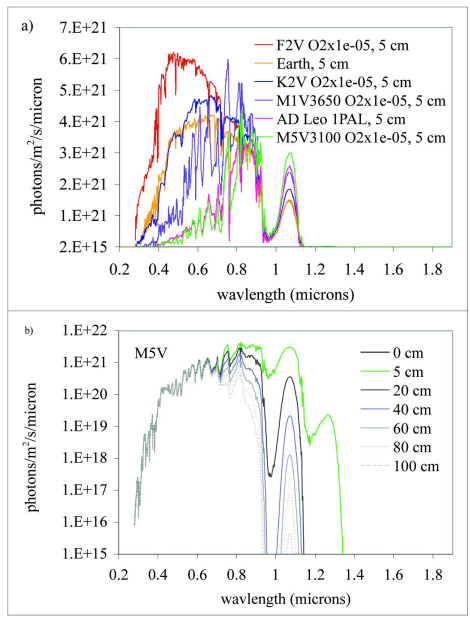

Figure 4. Underwater zenith photon flux densities for star-atmosphere scenarios in Table 2.
a) Photon fluxes at 5 cm depth in pure water, at solar noon at the equator.
b) Photon fluxes at the surface and several depths for a M5V 3100K star with negligible atmospheric O2.

**Figure 4 Underwater planetary maximum photon flux densities for star-atmosphere scenarios in Table 2. a) Photon fluxes at 5 cm depth in pure water. b) Photon fluxes at the surface and several depths for an M5V 3100K star with O2x10$^{-5}$ PAL.**